\documentclass[conference]{IEEEtran}
\IEEEoverridecommandlockouts
\usepackage{cite}
\usepackage{amsmath,amssymb,amsfonts}
\usepackage{algorithmic}
\usepackage{graphicx}
\usepackage{textcomp}
\def\BibTeX{{\rm B\kern-.05em{\sc i\kern-.025em b}\kern-.08em
    T\kern-.1667em\lower.7ex\hbox{E}\kern-.125emX}}

\usepackage[dvipsnames]{xcolor}
\usepackage{float}
\usepackage{threeparttable}
\usepackage{amssymb}
\usepackage{pifont}
\newcommand{\cmark}{\color{Emerald}{\ding{51}}}%
\newcommand{\xmark}{\color{Maroon}{\ding{55}}}%

\usepackage[ruled,linesnumbered,noend]{algorithm2e}
\usepackage{subfloat}
\usepackage{comment}
\usepackage[labelformat=simple]{subcaption}

\usepackage{multirow}
\usepackage{dblfloatfix}

\usepackage{arydshln}

\usepackage[T1]{fontenc}
\usepackage[most]{tcolorbox}
\usepackage{colortbl}
\usepackage{tikz}

\usepackage{listings}
\lstdefinelanguage{Verilog}{
  keywords={module, endmodule, input, output, wire, reg, always, if, else, assign, begin, end, property, assert, disable, iff, endproperty, posedge},
  keywordstyle=\color{blue}\bfseries,
  morecomment=[l]{//},
  morecomment=[s]{/*}{*/},
  commentstyle=\color{gray}\ttfamily,
  morestring=[b]",
  stringstyle=\color{red},
  identifierstyle=\color{black},
  basicstyle=\ttfamily\footnotesize,
  sensitive=true,
  showstringspaces=false,
  numbers=left,
  numberstyle=\tiny\color{gray},
  stepnumber=1,
  numbersep=5pt,
  breaklines=true,
  captionpos=b
}

\definecolor{codered}{rgb}{0.8,0,0}
\definecolor{codegreen}{rgb}{0,0.6,0}
\definecolor{codegray}{rgb}{0.5,0.5,0.5}
\definecolor{codepurple}{rgb}{0.58,0,0.82}
\definecolor{backcolour}{rgb}{0.95,0.95,0.92}
\definecolor{mycolor}{rgb}{0.90,0.95,0.90}

\lstdefinestyle{custom}{
    keywords={prompt_database},
    backgroundcolor=\color{backcolour},   
    commentstyle=\color{codegreen},
    keywordstyle=\color{blue}\bfseries,
    numberstyle=\tiny\color{codegray},
    stringstyle=\color{codepurple},
    basicstyle=\ttfamily\footnotesize,
    breakatwhitespace=false,         
    breaklines=true,                 
    captionpos=b,                    
    keepspaces=true,                 
    numbers=none,                    
    numbersep=5pt,                  
    showspaces=false,                
    showstringspaces=false,
    showtabs=false,                  
    tabsize=2,
    frame=tb 
}

\newcommand{\bstars}{\textsuperscript{$\bigstar$}}

\newcommand{\btrs}{\textsuperscript{$\blacktriangle$}}

\newcommand{\lasa}{{\texttt{LASA}}}

\begin{document}

\title{
LASA: Enhancing SoC Security Verification with LLM-Aided Property Generation
}

\author{
\IEEEauthorblockN{Dinesh Reddy Ankireddy, Sudipta Paria, Aritra Dasgupta, Sandip Ray, and Swarup Bhunia}
\IEEEauthorblockA{Department of Electrical and Computer Engineering, University of Florida, Gainesville, FL 32611, USA\\
\{ankired.dineshre, sudiptaparia, aritradasgupta\}@ufl.edu, \{sandip, swarup\}@ece.ufl.edu}
}

\maketitle

\begin{abstract}
Ensuring the security of modern System-on-Chip (SoC) designs poses significant challenges due to increasing complexity and distributed assets across the intellectual property (IP) blocks. Formal property verification (FPV) provides the capability to model and validate design behaviors through security properties with model checkers; however, current practices require significant manual efforts to create such properties, making them time-consuming, costly, and error-prone. The emergence of Large Language Models (LLMs) has showcased remarkable proficiency across diverse domains, including HDL code generation and verification tasks. Current LLM-based techniques often produce vacuous assertions and lack efficient prompt generation, comprehensive verification, and bug detection. This paper presents \lasa, a novel framework that leverages LLMs and retrieval-augmented generation (RAG) to produce non-vacuous security properties and SystemVerilog Assertions (SVA) from design specifications and related documentation for bus-based SoC designs. \lasa~integrates commercial EDA tool for FPV to generate coverage metrics and iteratively refines prompts through a feedback loop to enhance coverage. The effectiveness of \lasa~is validated through various open-source SoC designs, demonstrating high coverage values with an average of $\sim$88\% denoting comprehensive verification through efficient generation of security properties and SVAs. \lasa~also demonstrates bug detection capabilities, identifying five unique bugs in the buggy OpenTitan SoC from Hack@DAC'24 competition.
\end{abstract}


\section{Introduction}

Modern bus-based SoCs integrate multiple intellectual properties (IPs) on a single chip and utilizes a common bus to facilitate communication between them. With the globalization of the IC supply chain and adoption of a Zero Trust Model it becomes crucial to identify and fix vulnerabilities and protect secure assets in the regime of evolving security threats. Traditional verification methods including simulation, random-regression, directed-random testing \cite{book}, assertion-based formal verification \cite{formal_ver_survey_2}, fuzzing \cite{synfuzz}, and hybrid techniques \cite{fuzz_3} etc. struggle to keep pace with the growing complexity of SoC designs, highlighting the need for greater automation in vulnerability detection and remediation. Formal Property Verification (FPV) provides a rigorous methodology for mathematically verifying hardware design correctness using model-checking. However, the effectiveness of FPV depends on meticulously crafted security properties that accurately capture the intended behaviors and uncover potential vulnerabilities. Assertion Based Verification (ABV) uses assertions that are derived from its specification by the verification experts. These assertions are used to statically prove properties using formal verification tools or dynamically verified using simulation to identify potential vulnerabilities. Generating relevant security properties or assertions is a complex task involving substantial expertise and manual efforts by security experts, making it error-prone and not scalable for larger designs, highlighting the need for automation to streamline the verification process. 

The rapid evolution of Large Language Models (LLMs) has extended their capabilities beyond natural language processing, making a significant impact in the domain of hardware security and verification. LLMs have shown remarkable proficiency in automating tasks like HDL code generation, verification, and bug fixing \cite{llm_survey}. Recent studies on LLM-based generation of SystemVerilog Assertions (SVAs) \cite{orenesvera,kande,spell,assertllm}, security properties \cite{nspg,lasp} have demonstrated the growing potential of LLMs in automating such verification tasks. LLM-based bug fixing \cite{Pearce,llm_bug_fix_paper} aims to generate repairs for fixing security vulnerabilities involving static analysis and security-related feedback or policy-based enforcement \cite{divas}. However, these techniques are limited by the complexity of RTL designs and struggle to effectively incorporate diverse information sources, such as design specifications, threat models, and other security requirements, and lack in curating efficient prompts. Moreover, existing techniques lack vacuity checking, often resulting in the generation of vacuous or non-meaningful properties that hinder effective verification. These techniques also do not incorporate coverage analysis to assess the effectiveness of the generated properties, leaving ambiguity in terms of verification completeness. In this paper, we introduce a novel automated framework \lasa~(\underline{L}LM-\underline{A}ided \underline{S}ecurity Property Generation for \underline{A}ssertion-based SoC Verification) for efficiently generating security properties for comprehensive SoC verification and identifying bugs. 
This paper makes the following major contributions:
\begin{itemize}
    \item We propose \lasa, a novel and efficient framework that leverages the knowledge base of LLMs to automatically generate security properties and SVAs, enabling comprehensive verification of generic bus-based SoC designs.
    \item \lasa~integrates vacuity checking rules for identifying and discarding vacuous or non-meaningful properties to enhance verification efficiency and reduce the computational overhead. 
    \item \lasa~employs standard FPV tools to perform coverage analysis and incorporates an iterative refinement step when coverage falls below a threshold, ensuring improved and comprehensive verification.
    \item Experimental results demonstrate the effectiveness of \lasa~evaluated on open-source SoC benchmarks, achieving high coverage values that indicate substantial verification performance.
    \item \lasa~is equipped with bug detection capabilities, as evidenced by the identification of five bugs in the buggy OpenTitan benchmark from the HackDAC'24 competition.
\end{itemize}

The paper is organized as follows: Section II discusses the relevant background and related works. Section III presents major stages of the proposed framework. Section IV demonstrates the experimental results and discussion. Finally, Section V concludes the paper.

\section{Background}

\subsection{Security Property and Formal Property Verification}
Security property is a formal specification or rule describing an observable behavior that the hardware design must satisfy to guarantee confidentiality, integrity, and availability (CIA) requirements \cite{book,dispel}. These properties must adhere to three fundamental principles: Correctness, Consistency, and Completeness. Designers commonly use languages like PSL and SVAs, employing logic representations at the temporal level like LTL and CTL, to describe design behaviors. FPV is used to rigorously verify that security properties hold under all possible inputs. Commercial EDA tools employ model checkers to prove whether a property holds in all cases (safe) or find a counterexample (violation). By exhaustively proving or refuting properties, FPV enables security flaws detection that might evade traditional testing. 

\subsection{Vacuity Check Rules/Theorems for Security Properties}
\label{sec:vacuity_rule}

\noindent \textbf{Definition}: Vacuity in LTL refers to a situation where a specification (formula) is satisfied in a system (model), but not in a meaningful way as part of the formula was irrelevant in the system's behavior. Vacuous properties usually indicate weak, trivial, or incorrect specifications for a given system or design. 

\noindent \textbf{Example}:\\
Let us assume the following formula/property in Model $M$:
\setlength{\belowdisplayskip}{0pt} \setlength{\belowdisplayshortskip}{0pt}
\setlength{\abovedisplayskip}{0pt} \setlength{\abovedisplayshortskip}{0pt}
\begin{equation}
    \varphi = G(p \rightarrow Fq)
\end{equation}
This states ``Always, if $p$ holds, then eventually $q$ holds.''

\noindent In model $M$, if $p$ never occurs, then $\varphi$ is vacuously true because the implication 
$p \rightarrow Fq$ when $p$ is false.
Hence, we can conclude $\varphi$ is vacuously satisfied in $M$ with respect to $p$.

\noindent \textbf{Formal Definition:} (from \cite{vacuity_check,vacuity_check_2})\\
\noindent A system $M$ satisfies a formula $\varphi$ vacuously iff $M \models \varphi $ and there is some subformula $\psi$ of $\varphi$ such that $\psi$ does not affect $\varphi$ in $M$.
For example, verifying a system with respect to the specification $\varphi = AG(req \rightarrow AF grant)$ (``every request is eventually followed by a grant''), we say that $\varphi$ is satisfied vacuously in systems in which requests are never sent. \\

\noindent $\bullet$ \textbf{Theorems for Vacuity Checking\cite{vacuity_check}}

\noindent \textbf{Theorem 1. (Efficient vacuity checking)} For every formula $\varphi$, a subformula $\psi$ of $\varphi$, and a system $M$, the following are equivalent: \\
(1) $\psi$ does not affect $\varphi$ in $M$. \\
(2) $M$ satisfies $\varphi$[$\psi \leftarrow$ \textbf{\textit{true}}] iff $M$ satisfies $\varphi$[$\psi \leftarrow$ \textbf{\textit{false}}].

\noindent \textbf{Theorem 2. (Polynomial time complexity)} The problem of checking whether a system $M$ satisfies a formula $\varphi$ vacuously can be solved in time $\mathcal{O}(C_M (|\varphi|) \cdot |\varphi|)$.

\noindent \textbf{Theorem 3. (Complexity of checking vacuity in CTL)} For $\varphi$ in CTL,a subformula $\psi$ of $\varphi$ with multiple occurrences, and a system $M$, the problem of deciding whether $\psi$ does not affect $\varphi$ in $M$ is co-NP-complete.

\noindent \textbf{Theorem 4. (Linearly witnessable CTL formula)} Given a CTL formula $\varphi$, deciding whether $\varphi$ is linearly witnessable is in 2EXPTIME (Double Exponential Time or $\mathcal{O}(2^{2^{p(n)}}), \text{where~}p(n) \text{~denotes polynomial function of~}n$) and is EXPTIME(Exponential Time)-hard.

\noindent \textbf{Theorem 5. (Linearly counterable formula)} For a branching temporal logic formula $\varphi$, we have that $\varphi$ is linearly counterable iff $\neg \varphi$ is linearly witnessable. 

\noindent \textbf{Theorem 6. (Branching temporal logic)} For a branching temporal logic formula $\varphi$ and a system $M$, we have that $M \nvDash A \varphi^d$ iff $M$ has a path $\pi$ such that $\pi \nvDash \varphi$.

\noindent \textbf{Theorem 7. (Linearly Witnessable CTL$^\star$ formula)} For a CTL$^\star$ formula $\varphi$ and a system $M$, deciding whether $\varphi$ is linearly witnessable in $M$ is PSPACE-complete (polynomial amount of memory).

\noindent \textbf{Theorem 8. (Counterexamples to find interesting witnesses)} For a formula $\varphi$ and a system $M$, a counterexample for $\neg witness(\varphi)$ in $M$ is an interesting witness for $\varphi$ in $M$.

\noindent \textbf{Theorem 9. (Complexity of finding an interesting witness)} For an LTL or a CTL$^\star$ formula $\varphi$ and a system $M$, an interesting witness for $\varphi$ in $M$ can be generated in polynomial space. Deciding whether such a witness exists is PSPACE-complete.

\subsection{Coverage Metrics in FPV}

Commercial tools performing FPV offer a range of coverage metrics to assess the thoroughness and completeness of FPV. Cadence JasperGold employs three coverage metrics, namely, stimuli, checker, and formal coverage. 
\textbf{Stimuli coverage} indicates how well the input conditions and scenarios are applied to the DUT during formal verification. It includes both code coverage (such as branch, statement, expression, and toggle coverage) and functional coverage (defined by user-specified covergroups). \textbf{Checker Coverage} assesses the completeness of the formal assertions in verifying the design behavior. It includes \textit{Cone of Influence (COI)} Coverage which measures the extent to which the logic paths that influence a given assertion are exercised, and \textit{Proof Core} Coverage identifies the minimal set of design elements necessary for the assertion's truth value and ensures that these elements are thoroughly checked. \textbf{Formal Coverage} is a composite metric that combines both stimuli and checker coverage to provide an overall assessment of the formal verification's effectiveness.

\subsection{Existing Verification Techniques and Challenges}

The current industry practices predominantly employ two major methodologies for SoC verification: (i) simulation-based verification and (ii) assertion-based verification \cite{book,clip}. Simulation-based techniques rely on generating complex testbenches that drive inputs and monitor outputs under various scenarios. Tools such as ModelSim and Synopsys VCS are commonly employed to validate the functional correctness of designs. Simulation-based approach is widely used for its familiarity and ease of deployment, but it offers limited coverage and struggles with scalability as design complexity increases. ABV offers a more formal and rigorous framework that employs SVAs specifying intended design behavior through properties and constraints and verified using FPV. It provides exhaustive verification within bounded scopes and is particularly effective in uncovering subtle logic errors/bugs, or security vulnerabilities. However, it can be limited in scalability due to the state-explosion problem.

\setlength\dashlinedash{1.5pt}
\setlength\dashlinegap{3pt}
\setlength\arrayrulewidth{0.25pt}
\def\arraystretch{1.25}
\begin{table*}[!htbp]
\centering
\caption{Comparative analysis between \lasa~and existing LLM-based approaches.}
\label{tab:hdl_verification}
\resizebox{\textwidth}{!}{%
\begin{tabular}{|c|c|c|c|c|c|c|c|c|c|}
\hline
\cellcolor[HTML]{ECF4FF}\begin{tabular}[c]{@{}c@{}} \textbf{Proposed} \\ \textbf{Solutions} \end{tabular}             
& \cellcolor[HTML]{ECF4FF}\textbf{Baseline}            
& \cellcolor[HTML]{ECF4FF}\begin{tabular}[c]{@{}c@{}} \textbf{SVA} \\\textbf{Generation?} \end{tabular} 
& \cellcolor[HTML]{ECF4FF}\begin{tabular}[c]{@{}c@{}}\textbf{Property/Policy} \\ \textbf{Generation?} \end{tabular} 
& \cellcolor[HTML]{ECF4FF}\begin{tabular}[c]{@{}c@{}}\textbf{Vacuity}\\ \textbf{Checking?}\end{tabular} 
& \cellcolor[HTML]{ECF4FF}\begin{tabular}[c]{@{}c@{}}\textbf{Prompt} \\\textbf{Engineering?} \end{tabular} 
& \cellcolor[HTML]{ECF4FF}\textbf{RAG?} 
& \cellcolor[HTML]{ECF4FF}\begin{tabular}[c]{@{}c@{}} \textbf{Coverage} \\ \textbf{Analysis?} \end{tabular}
& \cellcolor[HTML]{ECF4FF}\begin{tabular}[c]{@{}c@{}}\textbf{Bug} \\\textbf{Detection?} \end{tabular} 
& \cellcolor[HTML]{ECF4FF}\begin{tabular}[c]{@{}c@{}}\textbf{Evaluation} \\\textbf{Benchmarks} \end{tabular} \\ 
\hline
Kande et al. \cite{kande}    
& Codex   
& \cmark
& \xmark
& \xmark
& \cmark
& \xmark
& \xmark
& \xmark
& Hack@DAC21, OpenTitan       
\\ 
NSPG \cite{nspg}                 
& BERT                                                            
& \xmark               
& \cmark                         
& \xmark          
& \cmark         
& \xmark 
& \xmark
& \xmark
& OpenTitan, RISC-V, MIPS       
\\ 
ChipNeMo \cite{chipnemo}           
& \begin{tabular}[c]{@{}c@{}}LLaMA2 7B/13B/70B \end{tabular}      
& \xmark               
& \xmark                           
& \xmark          
& \xmark      
& \cmark 
& \xmark
& \xmark
& Custom Benchmarks              
\\ 
AssertLLM \cite{assertllm}         
& \begin{tabular}[c]{@{}c@{}}GPT-3.5, GPT-4\end{tabular}    
& \cmark             
& \xmark                           
& \xmark            
& \cmark    
& \cmark 
& \xmark
& \xmark
& Custom Benchmarks             
\\ 
Hassan et al. \cite{hassan}     
& GPT-4                                                           
& \cmark             
& \xmark                           
& \xmark            
& \xmark          
& \xmark 
& \xmark
& \xmark
& C432 (ISCAS-85)         
\\ 
Saha et al. \cite{fics}             
&  GPT-3.5, GPT-4                                           
& \cmark             
& \xmark                           
& \xmark            
& \xmark           
& \xmark
& \xmark
& \xmark
& CWE, Trust-Hub        
\\ 
Orenes-Vera et al. \cite{orenesvera} 
& GPT-4                                               
& \cmark             
& \xmark                           
& \xmark            
& \cmark       
& \xmark 
& \xmark
& \xmark
& RISC-V CVA6 Ariane     
\\ 
SPELL \cite{spell}      
& GPT-3.5, GPT-4, Gemini 
& \cmark             
& \cmark                         
& \xmark            
& \cmark            
& \xmark 
& \xmark
& \xmark
& MIT-CEP SoC, CWE      
\\ 
\begin{tabular}[c]{@{}c@{}}LAAG RV \cite{laag_rv}\end{tabular}       
& GPT-4                                                                  
& \cmark         
& \xmark                   
& \xmark        
& \cmark         
& \xmark
& \xmark
& \xmark
& OpenTitan     
\\ 
FVEVAL \cite{FVEval}      
& GPT-4o, Gemini-1.5, Llama-3.1                                           
& \cmark            
& \xmark                        
& \xmark             
& \cmark      
& \xmark 
& \xmark
& \xmark
& Custom Benchmarks            
\\ 
LASP \cite{lasp}      
& Gemini-1.5                                             
& \cmark
& \cmark                         
& \xmark        
& \cmark      
& \xmark
& \xmark
& \xmark
& RSA, DES, SHA512, AES     
\\ 
\rowcolor[HTML]{dfd0ff}
\lasa* (This work)    
& GPT4o, Llama-3.1, Gemini-1.5                                          
& \cmark
& \cmark                         
& \cmark        
& \cmark      
& \cmark
& \cmark
& \cmark
& CEP, OpenTitan, Hack@DAC'24   
\\ \hline
\end{tabular}%
}
\vspace{-1.5em}
\end{table*}

\subsection{Related work on Leveraging LLMs in Verification}

The growing use of LLMs is greatly advancing state-of-the-art hardware security verification \cite{llm_survey} by leveraging their natural language understanding and broad knowledge base to perform diverse automation tasks. Recent research explores integrating LLMs into IC design flow to generate complex testbenches for simulation-based verification and creating precise, context-specific SVAs for formal verification. LLM-aided verification enables early detection of vulnerabilities prior to fabrication. LLM-based SVA generation \cite{kande,laag_rv,orenesvera,ip_cots} at the RTL abstraction level has been explored using natural language specifications and prompt engineering. Frameworks like AssertLLM \cite{assertllm} employ specialized LLMs for different stages of assertion creation, while hybrid methods \cite{hassan} combine LLM-driven formal verification with mutation testing to refine design invariants. ChipNeMo \cite{chipnemo} showcases versatile LLM applications, including chat-based assistance, script generation, and bug summarization, emphasizing domain-specific adaptation. DIVAS \cite{divas} provides an end-to-end toolflow that automates CWE identification, SVA generation, and security policy enforcement using \cite{dispel}, highlighting broad applications of LLMs. Additional frameworks such as FVEval \cite{FVEval}, LASP \cite{lasp}, and NSPG \cite{nspg} assess and generate security properties directly from RTL or design documentation. Table~\ref{tab:hdl_verification} summarizes the characteristics of existing solutions and also highlights the features of the proposed \lasa~framework.

\subsection{Motivation}

Current techniques face significant challenges due to the increasing complexity of SoC designs. They also lack the capability to curate efficient prompts by extracting relevant information from the corresponding documentation or design specifications. Current techniques follow more targeted verification and also do not guarantee the generation of non-vacuous properties that contribute to effective verification. Furthermore, existing approaches do not incorporate coverage analysis to evaluate the effectiveness of the generated properties or assertions, leaving gaps in verification completeness and increasing the risk of undetected errors. These limitations highlight the need for a more efficient automated framework that overcomes shortcomings of current approaches and ensures comprehensive verification of complex SoC designs.

\section{Methodology}

\begin{figure}[!ht]
    \centering
    \includegraphics[width=\columnwidth]{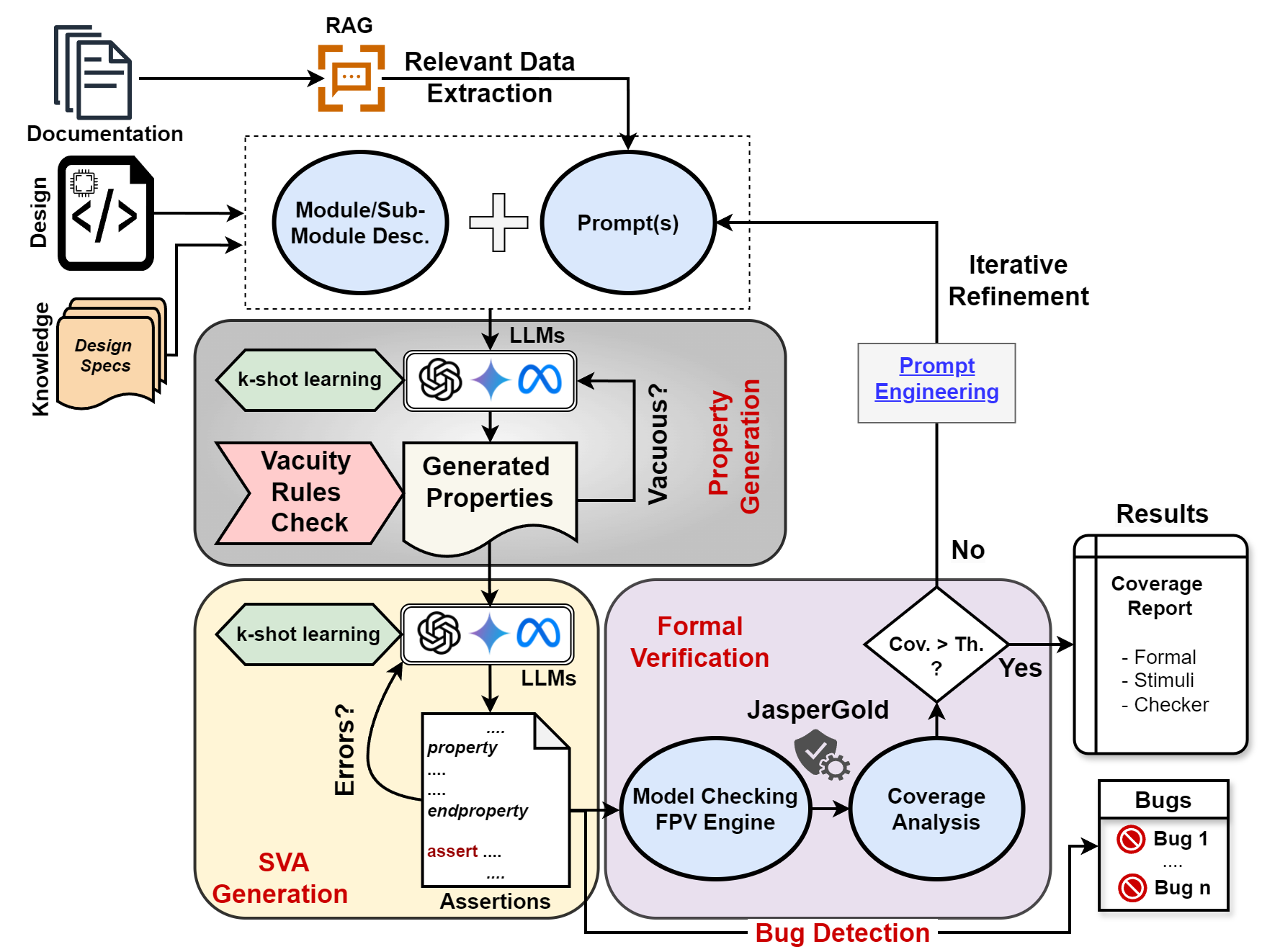}
    \caption{Major stages in the proposed \lasa~framework.}
    \label{fig:lasso_flow}
\end{figure}

In this section, we describe the main components of the \lasa~framework, starting from prompt creation, followed by non-vacuous security properties generation and then conversion to SVAs leveraging LLMs, along with coverage analysis with iterative refinement and bug detection. The overall flow has been depicted in Fig. \ref{fig:lasso_flow}. The proposed framework can be categorized into five major stages as described below.

\subsection{Prompt Generation}

The prompt generation step includes the extraction of behavioral descriptions from the given RTL code and specification documents, including block diagrams and textual descriptions. Current LLMs are constrained by the number of input tokens they can process, making it difficult for them to effectively analyze large hardware designs that consist of thousands of lines of RTL code. \lasa~breaks down the code into higher-level abstractions (e.g., modules, functions, or components) and uses the summarized descriptions in the creation of prompts. Additionally, \lasa~employs Retrieval-Augmented Generation (RAG) techniques to extract and summarize the relevant information from related documentation or manuals at a high level. By combining the summarized design descriptions and retrieval-based information, \lasa~generates more accurate, contextually aware prompts for generating the relevant security properties in the subsequent steps.

\subsection{Security Property Generation}

The proposed framework leverages pre-trained LLMs to generate relevant security properties for the SoC design under test. The security properties are typically expressed in temporal logic and explore specific or entire state space of the design, exploring possible states and transitions. The framework also incorporates vacuous rules checking to improve property generation and discard trivial or non-interesting properties. We borrow the standard nine vacuous rules and theorems (described in Section \ref{sec:vacuity_rule}) from literature \cite{vacuity_check,vacuity_check_2} in evaluating the vacuous properties. Additionally, the framework integrates k-shot learning for generating valid security properties while reducing hallucinations and irrelevant outputs from LLMs.

\subsection{SVA Generation}

\lasa~integrates pre-trained LLMs for translating non-vacuous security properties into equivalent SVAs. The framework employs re-prompting to correct syntax errors in generated assertions and produce revised versions. \lasa~appends each SVA into the respective .sva file and automates the process of generating TCL scripts for evaluation using Cadence JasperGold. The evaluation employs FPV using model checkers to generate counterexamples (CEXs) to identify any failures that infer potential bugs or vulnerabilities. \lasa~utilizes k-shot learning by allowing the model to learn from a small set of example assertions to generate syntactically accurate and contextually relevant SVAs. This allows \lasa~to adapt to diverse design specifications, producing high-quality assertions that align with the security goals and underlying threat model.

\subsection{Coverage Analysis}

Coverage analysis in FPV quantifies how effectively the verification process exercises the design under test and the generated SVAs derived from formal security properties. It provides quantitative metrics that help evaluate the completeness, effectiveness, and quality of the properties/assertions. \lasa~integrates multiple coverage metrics—Formal, Stimuli, and Checker Coverage—leveraging the Cadence JasperGold Coverage tool to analyze branch, statement, expression, and toggle coverage. \lasa~generates a comprehensive verification report summarizing assertion coverage and verification results. If any signals remain uncovered, further analysis can be performed by reviewing branch, statement, expression, toggle, and functional coverage reports.

\subsection{Iterative Refinement}

\lasa~employs iterative refinement that enhances the existing prompts to improve the coverage by generating additional properties followed by SVAs until the pre-defined coverage threshold (\%) is met. Failed SVAs indicate a potential issue—either due to incorrect formulation by LLM or uncovered design behavior that leads to potential bugs. The incorrect assertion generation by LLM necessitates prompt refinement through an iterative feedback path in \lasa~to enhance the assertion generation process and ensure greater accuracy. Additionally, \lasa~maintains additional prompts database and selectively applies relevant prompts aimed at generating more accurate properties, thereby enhancing coverage and improving overall verification completeness.

\section{Results and Discussion}

\subsection{Experimental set-up}
\begin{figure}[!ht]
\centering
\includegraphics[width=\columnwidth]{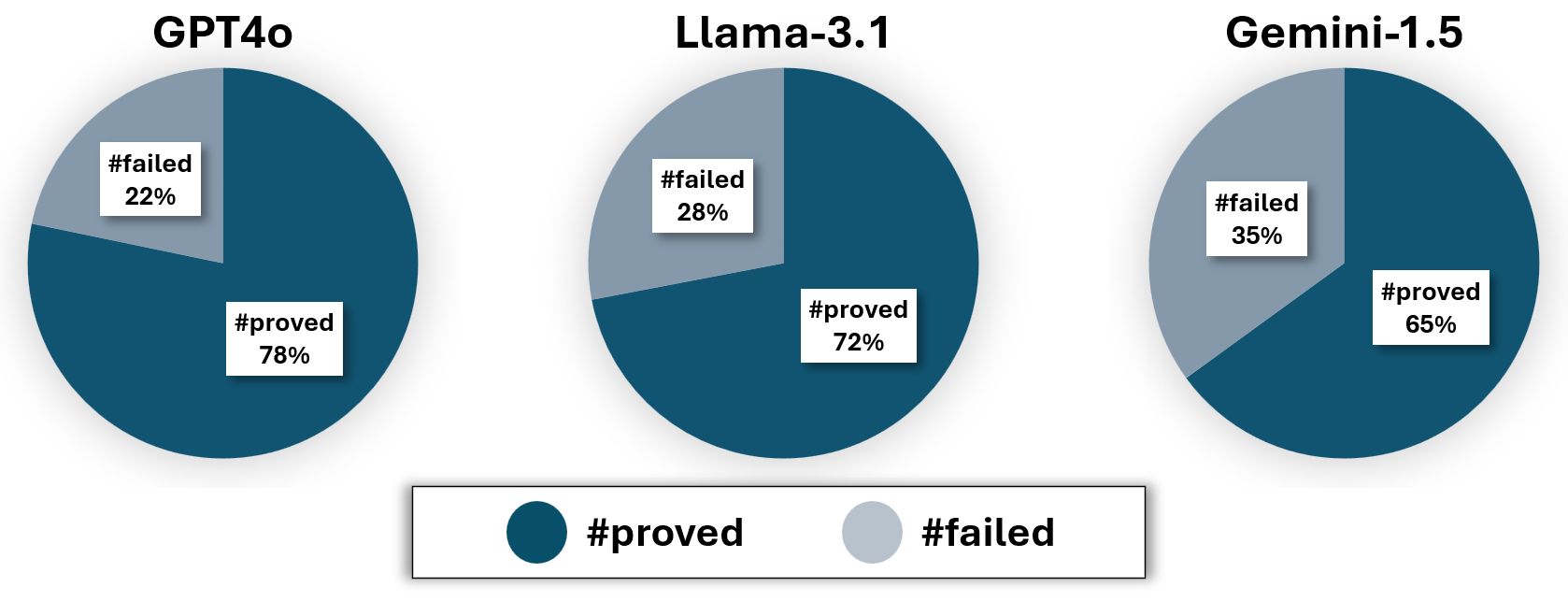}
\caption{Comparing LLMs in terms of generating properties.}
\label{fig:llm_comparison}
\end{figure}

We performed a comprehensive evaluation of our proposed \lasa~framework on two most popular open-source SoC benchmarks, namely, Common Evaluation Platform (CEP) from MIT-LL\footnote{https://github.com/mit-ll/CEP.git} and OpenTitan\footnote{https://opentitan.org/book/hw/ip/index.html}. The experiments were conducted on a Red Hat Enterprise Linux Server with AMD Epyc 7713 64-core processor and 1007.6 GiB Memory. FPV has been performed using Cadence JasperGold (ver. 2020.12). Our analysis utilizes diverse pre-trained LLMs with their latest available versions such as OpenAI's GPT4o, Google’s Gemini-1.5, and Meta's Llama-3.1 for generating properties and SVAs. We evaluated the performance of different pre-trained LLMs and found GPT4o outperforms other LLMs based on the percentage of generated security properties classified as \#proved or \#failed averaged on multiple IPs (refer to Fig. \ref{fig:llm_comparison}). Hence, we selected GPT-4o as our primary LLM for integration with \lasa~for automation tasks in different stages.

\subsection{Generation of Initial Prompts and RAG-based Extraction}

\lasa~adopts JSON-based template (<SPEC\_FILE>) as shown partially in Listing \ref{lst:2} that formalize the specifications. For complex SoCs with multiple IPs involved, \lasa~processes each IP individually and uses the corresponding IP-level documentation and <SPEC\_FILE> to generate tailored prompts. For large hierarchical IPs, \lasa~identifies the submodules based on design specifications and performs evaluation at the submodule level. The initial prompt includes high-level design specifications extracted from <SPEC\_FILE>, either at the module or submodule level, augmented with contextual information retrieved via RAG-based techniques. These enriched prompts are fed sequentially to pre-trained LLMs to facilitate the generation of relevant security properties while mitigating the token limitations of LLMs. 
The example prompts are included in Appendix \ref{appendix:prompts}.

\lstset{style=custom}
\begin{lstlisting}[caption=CEP SoC Design Specifications in JSON format., label=lst:2] 
"SoC_General":
{
    "NAME":"MIT-CEP",
    "TYPE":"Open-source",
    "BUS":"AXI4",
    "NO_OF_IP":"12",
    .... // more details
},
"BUS_INTERFACE":
{
    "INTERFACE_NAME":"Master/Slave",
    "NO_OF_PORTS":"17",
    .... // more details
}
"IP_1":
{
    "NAME":"AES",
    "TYPE":"Slave",
    "OPERATION":"Crypto",
    .... // more details
}
.... // more IPs
"Assets":
{
    "NAME":"aes_key",
    "TYPE":"192-bit",
    .... // more details
}
\end{lstlisting}

\subsection{Generation of Security Properties and Vacuity Check}

\begin{figure}[!ht]
\centering
\includegraphics[width=0.9\columnwidth]{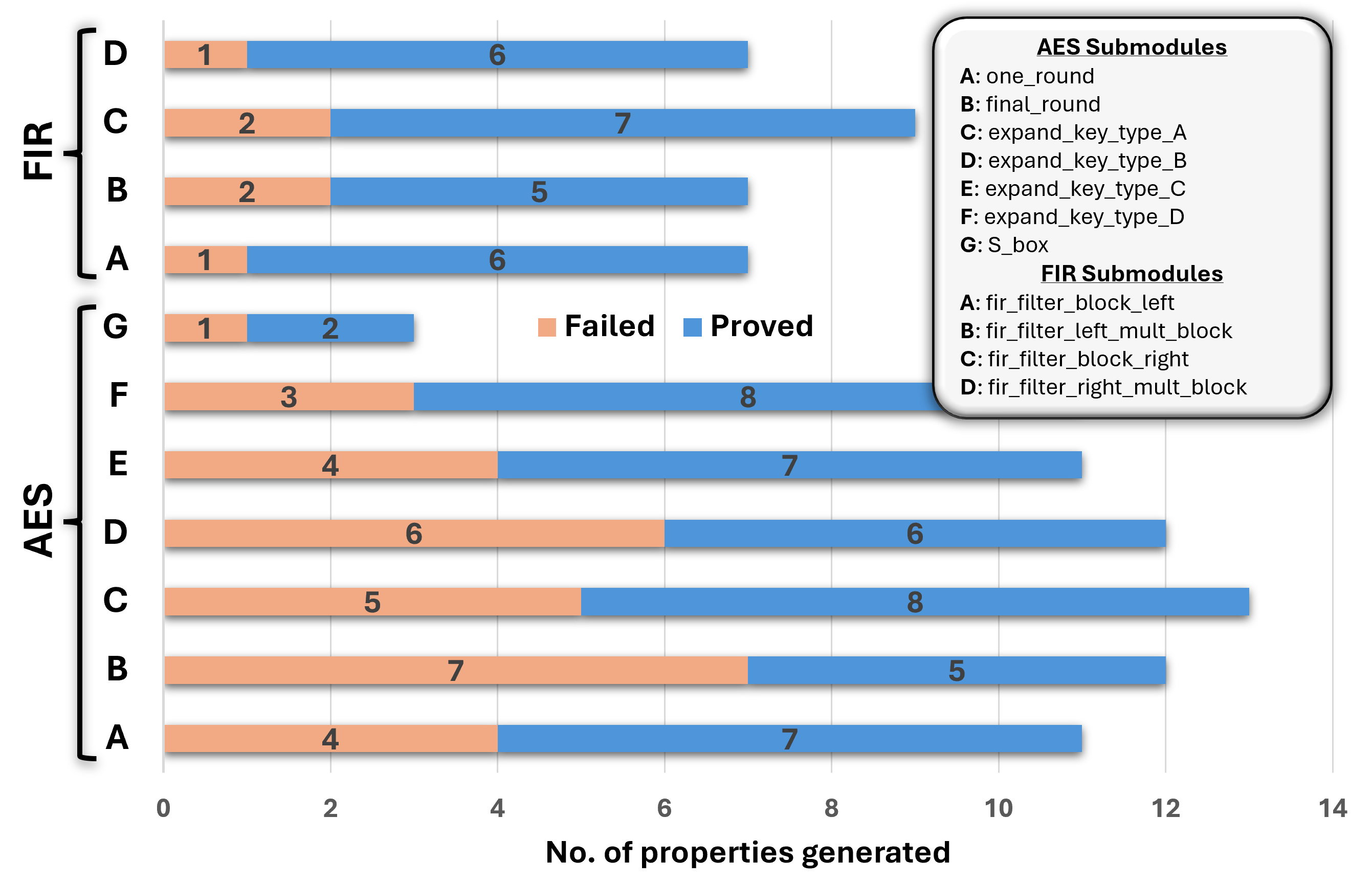}
\caption{Number of `Proved' and `Failed' Properties Generated by \lasa~for AES and FIR Sub-Modules.}
\label{fig:submodule_level_data}
\end{figure}

\lasa~employs vacuity check on the LLM-generated properties to discard all non-vacuous properties that violate at least one of nine theorems (Section \ref{sec:vacuity_rule}). 
\lasa~adopts a hierarchical analysis approach, treating each submodule independently to generate localized security properties, which are subsequently aggregated at the top module level to ensure comprehensive coverage and consistency.  
Table \ref{tab:performance} presents the number of security properties generated for different IPs from CEP and OpenTitan benchmarks. The column `\#correct' denotes the number of properties that are proven to be relevant and semantically correct, and `\#non-vacuous' denotes the number of properties that pass vacuity check rules. Fig. \ref{fig:submodule_level_data} represents the number of proved and failed properties generated for each submodule of \texttt{AES} and \texttt{FIR} designs from CEP.

\begin{table}[!htbp]
\centering
\caption{Number of generated non-vacuous properties and correct SVAs using \lasa.}
\label{tab:performance}
\resizebox{\columnwidth}{!}{%
\begin{tabular}{|c|ccc|cc|}
\hline
\multicolumn{1}{|c|}{\multirow{2}{*}{\textbf{Design}}} &
  \multicolumn{3}{c|}{\cellcolor[HTML]{ECF4FF}\textbf{Properties}} &
  \multicolumn{2}{c|}{\cellcolor[HTML]{ECF4FF}\textbf{Assertions}} \\ \cline{2-6} 
\multicolumn{1}{|c|}{} &
  \multicolumn{1}{c|}{\cellcolor[HTML]{ECF4FF}\textbf{\#generated}} &
  \multicolumn{1}{c|}{\cellcolor[HTML]{ECF4FF}\textbf{\#correct}} &
  \multicolumn{1}{c|}{\cellcolor[HTML]{ECF4FF}\textbf{\#non-vacuous}} &
  \multicolumn{1}{c|}{\cellcolor[HTML]{ECF4FF}\textbf{\#generated}} &
  \multicolumn{1}{c|}{\cellcolor[HTML]{ECF4FF}\textbf{\#correct}} \\ \hline
 
 AES-192\btrs & \multicolumn{1}{c}{73} & 43 & 41 & \multicolumn{1}{c}{41} & 40 \\ \hline
 DES3\btrs    & \multicolumn{1}{c}{37} & 31 & 29 & \multicolumn{1}{c}{29} & 27 \\ \hline
 GPS\btrs    & \multicolumn{1}{c}{108} & 98 & 93 & \multicolumn{1}{c}{93} & 90 \\ \hline
 FIR\btrs    & \multicolumn{1}{c}{30} & 24 & 22 & \multicolumn{1}{c}{22} & 22 \\ \hline
 IIR\btrs    & \multicolumn{1}{c}{38} & 26 & 25 & \multicolumn{1}{c}{25} & 25 \\ \hline
 i2c\bstars    & \multicolumn{1}{c}{78} & 75 & 71 & \multicolumn{1}{c}{71} & 70 \\ \hline
 adc\_ctrl\bstars    & \multicolumn{1}{c}{37} & 33 & 32 & \multicolumn{1}{c}{32} & 32 \\ \hline
 kmac\bstars    & \multicolumn{1}{c}{49} & 47 & 44 & \multicolumn{1}{c}{44} & 42 \\ \hline
\end{tabular}%
}
\footnotesize
{\vspace{0.75mm} \btrs denotes designs from CEP SoC. \hspace{0.5em} \bstars denotes designs from OpenTitan SoC.}\raggedright
\end{table}

\subsection{Generation of Valid SVAs}

\lasa~leverages pre-trained LLMs to generate equivalent assertions for all non-vacuous properties that pass the vacuity check. We found most of the properties were directly translatable into SVAs (refer to Table \ref{tab:performance}) while some posed challenges for direct conversion due to the dependence on the formulation of properties. There were some manual efforts involved for those properties requiring re-prompting for conversion to SVAs. LLMs often struggle with generating correct SVAs due to limited understanding of hardware behavior, signal timing, and implication semantics (e.g., overlapping vs. non-overlapping), etc. LLMs may generate SVAs that are syntactically valid but semantically incorrect such as using non-boolean expressions in implication, incorrect signal scopes, or violating reset behavior constraints. \lasa~integrates verification through Cadence JasperGold and discards the syntactically incorrect or semantically invalid SVAs. 

\noindent \textbf{Example of Vacuous Property/Assertion:}\\
Listing \ref{lst:vacuous} demonstrates an example property that violates one of the nine theorems of vacuity checking and will be treated as a vacuous property, and the corresponding assertion would also be discarded. This property is potentially vacuous since the antecedent  
\texttt{validCounter $>$ 1} is never \texttt{true} if \texttt{validCounter} is stuck at 1 or 0.

\begin{lstlisting}[language=Verilog, caption={Example of Vacuous Property/Assertion.}, label=lst:vacuous]
property p_validCounter_decrement;
  @(posedge clk) disable iff (rst)
  (validCounter > 1) |=> (validCounter == $past(validCounter) - 1);
endproperty
assert property (p_validCounter_decrement);
\end{lstlisting}

\begin{figure}[!ht]
\centering
\includegraphics[width=\columnwidth]{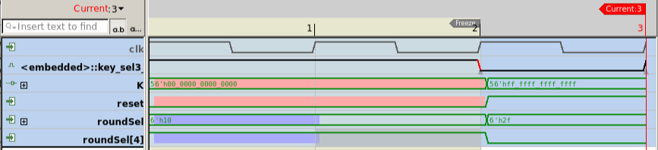}
\caption{Timing diagram depicting the counterexample (CEX).}
\label{fig:counterexample}
\end{figure}

\noindent \textbf{Example of CEX:}\\
Listing \ref{lst:vacuous} illustrates a counterexample generated for an LLM-generated SVA for \texttt{DES3} design. This condition violates the design behavior, and model checking in FPV can generate a counterexample for the same (refer to Fig. \ref{fig:counterexample}), making it semantically incorrect.

\begin{lstlisting}[language=Verilog, caption={Example of CEX for DES3 design.}, label=lst:vacuous]
property p_k_update;
  @(posedge clk) disable iff (reset)
  (!$stable(roundSel)) |=> (K !== $past(K));
endproperty
assert property (p_k_update);
\end{lstlisting}

\vspace{-1em}

\subsection{Coverage Analysis}

\lasa~integrates Cadence JasperGold to perform FPV on the generated SVAs in the previous step and generate respective coverage metrics. \lasa~follows a configurable coverage threshold (set to 80\% for our experiments), representing the minimum coverage required for the verification to be considered \textit{acceptable}. If coverage falls below this threshold, \lasa~incoprates a feedback loop to generate additional relevant SVAs to improve the coverage. Higher coverage \% indicates a more thorough and efficient verification, ensuring that all relevant states, behaviors, and properties of the design are fully explored and validated. \lasa~can also be seamlessly integrated with other commercial EDA tool flows to perform FPV and generate respective coverage metrics. Table \ref{tab:coverage} presents the highest coverage values achieved for eight different IPs belonging to CEP and OpenTitan SoC. Coverage values at sub-module level are detailed in Appendix \ref{appendix:cov_submodule}.

\begin{table}[!htbp]
\centering
\caption{Coverage values (\%) for different IPs.}
\label{tab:coverage}
\resizebox{\columnwidth}{!}{%
\begin{tabular}{|c|cc|c|cc|}
\hline
\multirow{2}{*}{\textbf{Benchmarks}} &
  \multicolumn{2}{c|}{\textbf{\cellcolor[HTML]{ECF4FF}Checker Coverage}} &
  \multirow{2}{*}{\textbf{\begin{tabular}[c]{@{}c@{}}\cellcolor[HTML]{ECF4FF}Stimuli \\ \cellcolor[HTML]{ECF4FF}Coverage\end{tabular}}} &
  \multicolumn{2}{c|}{\textbf{\cellcolor[HTML]{ECF4FF}Formal Coverage}} \\ \cline{2-3} \cline{5-6} 
 &
  \multicolumn{1}{c|}{\textbf{\cellcolor[HTML]{ECF4FF}COI}} &
  \textbf{\cellcolor[HTML]{ECF4FF}Proof Core} &
   &
  \multicolumn{1}{c|}{\textbf{\cellcolor[HTML]{ECF4FF}COI}} &
  \textbf{\cellcolor[HTML]{ECF4FF}Proof Core} \\ \hline
  \multicolumn{1}{|c|}{AES-192\btrs} &
  \multicolumn{1}{c|}{99.02\%} & 83.59\%
   & \multicolumn{1}{c|}{99.93\%} &
  \multicolumn{1}{c|}{99.37\%} & 85.05\%
   \\ \hline    
   \multicolumn{1}{|c|}{DES3\btrs} &
  \multicolumn{1}{c|}{93.97\%} & 91.70\%
   & \multicolumn{1}{c|}{100\%} &
  \multicolumn{1}{c|}{93.98\%} & 91.71\%
   \\ \hline  
   \multicolumn{1}{|c|}{GPS\btrs} &
  \multicolumn{1}{c|}{98.48\%} & 83.39\%
   & \multicolumn{1}{c|}{99.40\%} &
  \multicolumn{1}{c|}{98.63\%} & 84.70\%
   \\ \hline  
   \multicolumn{1}{|c|}{FIR\btrs} &
  \multicolumn{1}{c|}{92.64\%} & 80.35\%
   & \multicolumn{1}{c|}{92.82\%} &
  \multicolumn{1}{c|}{91.59\%} & 80.50\%
   \\ \hline  
   \multicolumn{1}{|c|}{IIR\btrs} &
  \multicolumn{1}{c|}{90.68\%} & 81.07\%
   & \multicolumn{1}{c|}{99.01\%} &
  \multicolumn{1}{c|}{90.97\%} & 81.42\%
   \\ \hline  
   \multicolumn{1}{|c|}{i2c\bstars} &
  \multicolumn{1}{c|}{91.44\%} & 84.40\%
   & \multicolumn{1}{c|}{99.25\%} &
  \multicolumn{1}{c|}{91.44\%} & 84.40\%
   \\ \hline  
   \multicolumn{1}{|c|}{adc\_ctrl\bstars} &
  \multicolumn{1}{c|}{86.37\%} & 80.53\%
   & \multicolumn{1}{c|}{95.18\%} &
  \multicolumn{1}{c|}{86.12\%} & 80.78\%
   \\ \hline  
   \multicolumn{1}{|c|}{kmac\bstars} &
  \multicolumn{1}{c|}{89.35\%} & 81.53\%
   & \multicolumn{1}{c|}{96.79\%} &
  \multicolumn{1}{c|}{90.22\%} & 82.19\%
   \\ \hline  
   \rowcolor{gray!20}
   \multicolumn{1}{|c|}{\textbf{Average}} &
  \multicolumn{1}{c|}{\%} & \%
   & \multicolumn{1}{c|}{\%} &
  \multicolumn{1}{c|}{\%} & \%
   \\ \hline  
\end{tabular}%
}
\footnotesize
{\vspace{0.75mm} \btrs denotes designs from CEP SoC. \hspace{0.5em} \bstars denotes designs from OpenTitan SoC.}\raggedright
\vspace{-1em}
\end{table}

\begin{table*}[!ht]
\centering
\caption{Detected bugs from HackDAC'24 buggy OpenTitan SoC using \lasa~framework.}
\label{tab:bug_detection}
\resizebox{\textwidth}{!}{%
\begin{tabular}{|c|c|c|c|l|}
\hline
\textbf{Bug\#} &
  \textbf{\begin{tabular}[c]{@{}c@{}}Bug Description\end{tabular}} &
  \textbf{\begin{tabular}[c]{@{}c@{}} Code Reference\end{tabular}} &
  \textbf{\begin{tabular}[c]{@{}c@{}}Security impact\end{tabular}} &
  \multicolumn{1}{c|}{\textbf{SVA for Detection}} \\ \hline
1 &
  \begin{tabular}[c]{@{}c@{}}Incorrect   parity checks for UART receiver. \\ The rx\_parity\_err in uart\_rx does not depend \\ on parity\_enable\end{tabular} &
  \begin{tabular}[c]{@{}c@{}}uart\_rx.sv:   \\ Line: 102-103\end{tabular} &
  \begin{tabular}[c]{@{}c@{}}It can lead to incorrect FIFO sync \\ (prim\_fifo\_sync) and false hardware \\ interrupt   (intr\_hw\_rx\_parity\_err).\end{tabular} &
  \begin{tabular}[c]{@{}l@{}} property p\_parity\_err\_without\_enable;\\        @(posedge clk\_i) disable iff   (!rst\_ni)\\        (!parity\_enable \&\&   rx\_valid\_q) |-\textgreater !rx\_parity\_err;\\      endproperty\\     assert property(p\_parity\_err\_without\_enable);\end{tabular} \\ \hline

2 &
  \begin{tabular}[c]{@{}c@{}}Incorrect wr\_data is assigned. Assigned q \\ (reg value) instead of d (hw value)\end{tabular} &
  \begin{tabular}[c]{@{}c@{}}prim\_subreg\_arb.sv: \\ Line: 51\end{tabular} &
  \begin{tabular}[c]{@{}c@{}}Incorrect assignment leads to \\ non-clearance of hw value.\end{tabular} &
  \begin{tabular}[c]{@{}l@{}}property p\_w1s\_hw\_clear\_should\_use\_d;\\   @(posedge clk\_i) disable iff (!rst\_ni)\\   (SwAccess == SwAccessW1S \&\& de \&\& !we)\\   |-\textgreater (wr\_data === d); \\ endproperty\\ assert property(p\_w1s\_hw\_clear\_should\_use\_d);\\   \end{tabular} \\ \hline

3 &
  \begin{tabular}[c]{@{}c@{}}Potential OTP word Overflow\end{tabular} &
  \begin{tabular}[c]{@{}c@{}}otp\_ctrl\_lci.sv:\\Line: 67\end{tabular} &
  \begin{tabular}[c]{@{}c@{}}Overflow will cause system interruption\\ by generating wrong values.\end{tabular} &
  \begin{tabular}[c]{@{}l@{}}property p\_otp\_word\_check;\\
    @(posedge clk\_i) disable iff (!rst\_ni)\\
    (LastLcOtpWord != LastLcOtpWordInt[CntWidth-1:0]);\\
endproperty\\
assert property(p\_otp\_word\_check);\\   \end{tabular} \\ \hline

4 &
  \begin{tabular}[c]{@{}c@{}}HMAC hashing key leaked through \\ reg\_rdata\_next\end{tabular} &
  \begin{tabular}[c]{@{}c@{}}hmac\_reg\_top.sv:   \\ Line: 1267-1273; 1343-1345;\end{tabular} &
  \begin{tabular}[c]{@{}c@{}}Malicious   attempt to leak the HMAC \\ hashing key.\end{tabular} &
  \begin{tabular}[c]{@{}l@{}}property   p\_hmac\_key\_read\_blocked;\\  @(posedge clk\_i)\\    disable iff (!rst\_ni)\\    (addr\_hit{[}8{]} || addr\_hit{[}9{]}) |-\textgreater reg\_rdata\_next == '0;\\ endproperty\\ assert property(p\_hmac\_key\_read\_blocked);\end{tabular} \\ \hline
5 &
  Incorrect error detection logic &
  \begin{tabular}[c]{@{}c@{}}prim\_subreg\_shadow.sv:\\Line: 76-66; 184-185;\end{tabular} &
  \begin{tabular}[c]{@{}c@{}}Since   error\_s is not used hence the error \\ detection logic will not work correctly   \\ leading to incorrect operation.\end{tabular} &
  \begin{tabular}[c]{@{}l@{}}property p\_error\_s\_known;\\    @(posedge clk\_i)\\     disable iff (!rst\_ni)\\      \$isunknown(error\_s) == 0;\\  endproperty\\   assert property(p\_error\_s\_known);\end{tabular} \\ \hline

\end{tabular}%
}
\end{table*}

\subsection{Iterative Refinement}

\begin{figure}[!ht]
\centering
\includegraphics[width=\columnwidth]{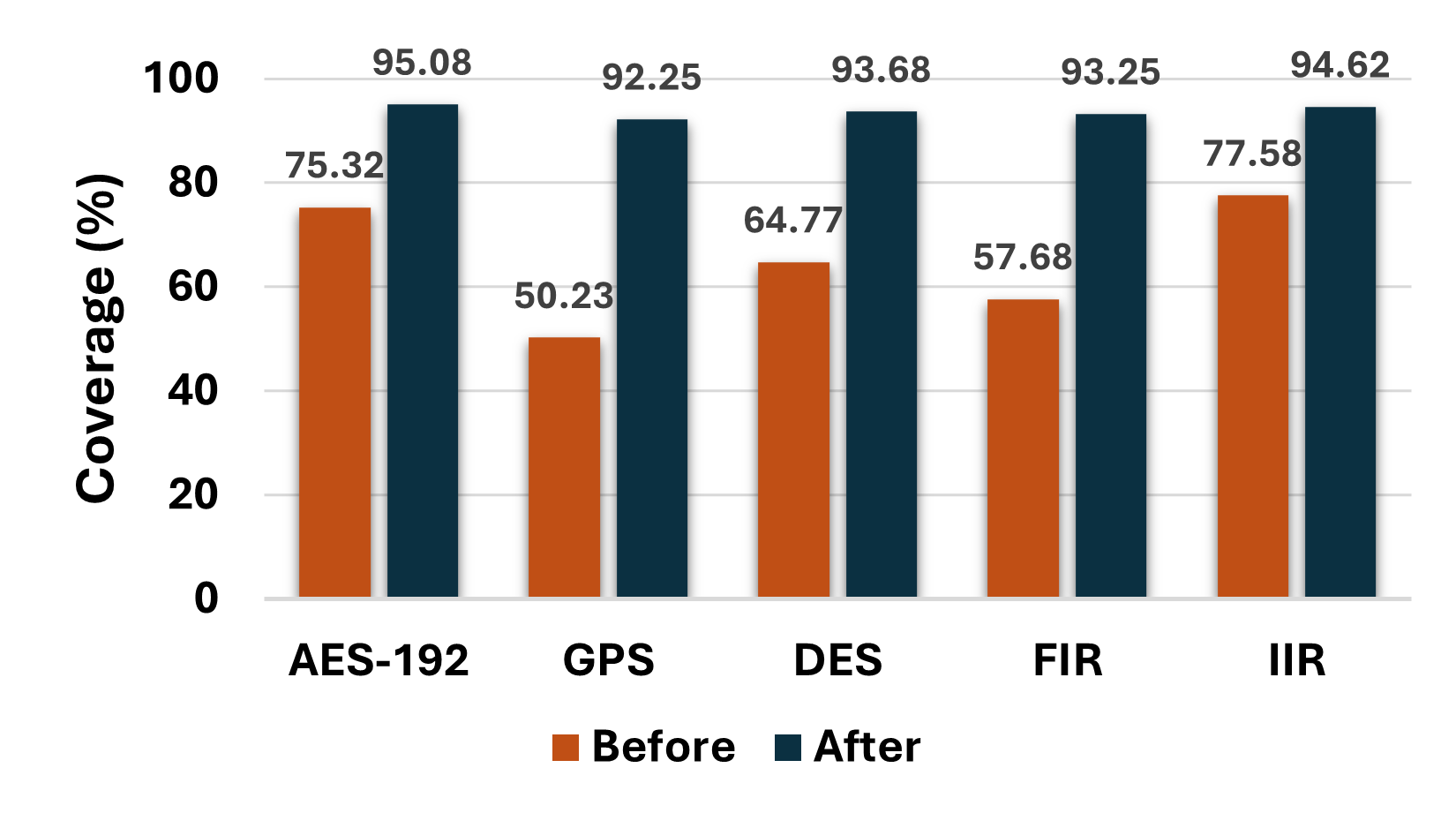}
\caption{Improvement on coverage values for different IPs through iterative refinement in \lasa.}
\label{fig:iterative_improvement}
\end{figure}

\lasa~includes a feedback path involving prompt engineering to regenerate more non-vacuous properties and context-specific SVAs for further improvement of coverage metrics if it falls below the threshold. Listing \ref{lst:prompt} shows some example prompts providing additional semantic context that are included with the existing prompts, refining the requirements and generating relevant properties and SVAs. Fig. \ref{fig:iterative_improvement} illustrates the iterative improvement on coverage metrics across five different designs from the CEP SoC benchmark. This iterative approach ensures that \lasa~enhances formal coverage, leading to more robust and comprehensive verification outcomes.

\vspace{1em}
\begin{lstlisting}[caption={Example prompts for iterative refinement.}, label=lst:prompt]
prompt_database:
  "Enhance coverage by adding reset conditions."
  "Add corner case assertions for boundary values."
  "Include sequential behavior checks."
  "Cover unreachable states if any."
  "Extend assertions to multi-cycle paths."
\end{lstlisting}

\subsection{Bug Detection}
\lasa~facilitates bug detection by generating context-aware SVAs that capture the security and functional properties of the design. For generating tailored SVAs for detecting bugs, \lasa~includes additional high-level information into the prompts, including micro-architectural events, threat models, secure assets, etc. These improved prompts guide the language model to produce tailored SVAs that more accurately reflect the design's security objectives. When verified, these SVAs can expose violations that indicate design flaws. This capability is demonstrated through the detection of five bugs in the buggy OpenTitan SoC from the Hack@DAC'24 competition and tabulated in Table \ref{tab:bug_detection}.

\subsection{Discussion}

While \lasa~demonstrates promising results, further enhancements are still possible. Incorporating more robust vacuity checking rules or refining existing ones could help filter out vacuous properties. LLMs struggle with complex, specialized tasks and are constrained by a finite set of use cases in generating SVAs, often leading to incomplete or erroneous responses. Fine-tuning LLMs on domain-specific datasets can significantly enhance their performance. Integrating syntax-checking tools helps minimize errors in generated responses and significantly reduces manual efforts. Enhancing LLMs with advanced reasoning capabilities and more sophisticated prompt engineering could improve the automation tasks. Additionally, adopting custom coverage metrics would offer a more accurate representation of verification completeness, guiding iterative improvements effectively.

\section{Conclusion}

In this paper, we presented \lasa, a novel efficient framework for automating security verification for generic bus-based SoCs. \lasa~incorporates RAG-based implementation to extract relevant data from hardware documentation and formalize SoC specifications for generating effective LLM prompts. By leveraging pre-trained LLMs and combining vacuity checking and k-shot learning, \lasa~efficiently generates contextually relevant non-vacuous security properties and then SVAs at both module and sub-module levels. The experimental results demonstrate higher coverage values (avg. 88\%) with iterative refinement, highlighting \lasa's effectiveness in comprehensive verification. \lasa~also showcases bug detection capabilities, enabling the identification of potential vulnerabilities at the early stages of the design flow. Future work includes incorporating domain-adapted LLMs with fine-tuning to improve the performance and also extending to other SoC interconnect fabrics, e.g., Network-on-Chip (NoC).

\bibliographystyle{IEEEtran}
\bibliography{IEEEabrv,references}

\appendix

\subsection{Example Prompts}
\label{appendix:prompts}

\begin{tcolorbox}[colback=gray!5!white,colframe=black!50!black,title=Prompt 1 : Initial Prompt]
You are an expert in Formal Verification and in generating SystemVerilog Assertions, specializing in writing comprehensive properties, including liveness, safety, and fairness constraints. You ensure thorough coverage by checking for deadlock, livelock, and starvation scenarios. To manage state space explosion, you effectively apply constraints, symbolic tokens, and assumptions. You also utilize auxiliary code and strive to write efficient, modular properties in place of overly complex assertions instead for Hardware Designs.
\end{tcolorbox}

\begin{tcolorbox}[colback=gray!5!white,colframe=black!50!black,title=Prompt 2 : Vacuity checking rules for k-shot learning]
-- // 9 Rules <give in the text file> or direct in prompt 
Here are the 9 vacuity checking rules you need to memorize. Utilize these 9 theorems to generate valid and non-vacuous properties for the <module/sub-module>.
\end{tcolorbox}

\begin{tcolorbox}[colback=gray!5!white,colframe=black!50!black,title=Prompt 3: Vacuity checking of generated properties]
Can you evaluate and check if each property generated is non-vacuous? A property must fail at least one vacuity condition among the 9 theorems to be non-vacuous. Property is vacuous or non-vacuous using the 9 vacuity rules provided, and respond as if Non-Vacuous True else False. 
\end{tcolorbox}

\begin{tcolorbox}[colback=gray!5!white,colframe=black!50!black,title=Prompt 4: Generating Counter Examples (CEXs)]
Can you help analyze the Counter Examples (CEXs) report from formal verification runs. Please analyze and explain the failure in the given CEXs and generate or modify the corresponding Assertions.
\end{tcolorbox}

\begin{tcolorbox}[colback=gray!5!white,colframe=black!50!black,title=Prompt 5: Fixing SVA errors]
There is an error in the properties generated by you. Check for Syntax or Semantics errors and correct it accordingly. Here is the error
[ERROR (error code: VERI-1137)] <.sva> syntax error near <>
\end{tcolorbox}

\begin{tcolorbox}[colback=gray!5!white,colframe=black!50!black,title=Prompt 6: Generating corrected SVAs through iterative refinement]
Based on this coverage report <statement, toggle, expression, branch> analysis, so the following cases <?> are not covered, can you generate more property-based system verilog assertions to cover all the cases.
\end{tcolorbox}

\begin{tcolorbox}[colback=gray!5!white,colframe=black!50!black,title=Prompt 7: Design specs.]
You need to generate Properties for the hardware verification of some designs. Here is the Hardware specification <SPEC FILE>. Based on this, you need to use the RAG model to prepare a module description at a high level such that I need to give the described design specification to the LLM model to generate some properties-based assertions. For each signal, extract the following information:
1. Signal name
2. Port Declearation:-Output, Input
3. Description: Definition, bit width, Signal type
4. Functionality
5. Any interconnects with other signals
6. Additional information required for assertions
7. Microarchitecture design
\end{tcolorbox}

\begin{table*}[b]
\centering
\caption{Coverage values for different IPs at sub-module level.}
\label{tab:cov_sub}
\resizebox{0.8\textwidth}{!}{%
\begin{tabular}{|c|c|cc|c|cc|}
\hline
\rowcolor[HTML]{FFFFFF} 
\cellcolor[HTML]{FFFFFF} &
  \cellcolor[HTML]{FFFFFF} &
  \multicolumn{2}{c|}{\cellcolor[HTML]{FFFFFF}\textbf{Checker   Coverage Settings}} &
  \cellcolor[HTML]{FFFFFF} &
  \multicolumn{2}{c|}{\cellcolor[HTML]{FFFFFF}\textbf{Formal   Coverage}} \\ \cline{3-4} \cline{6-7} 
\rowcolor[HTML]{FFFFFF} 
\multirow{-2}{*}{\cellcolor[HTML]{FFFFFF}\textbf{Benchmarks}} &
  \multirow{-2}{*}{\cellcolor[HTML]{FFFFFF}\textbf{Sub module}} &
  \multicolumn{1}{c|}{\cellcolor[HTML]{FFFFFF}\textbf{COI}} &
  \textbf{Proof Core} &
  \multirow{-2}{*}{\cellcolor[HTML]{FFFFFF}\textbf{Stimuli Coverage}} &
  \multicolumn{1}{c|}{\cellcolor[HTML]{FFFFFF}\textbf{COI}} &
  \textbf{Proof Core} \\ \hline
\cellcolor[HTML]{FFFFFF} &
  AES\_192\_top &
  \multicolumn{1}{c|}{97.94\%} &
  81.25\% &
  99.56\% &
  \multicolumn{1}{c|}{97.52\%} &
  85.50\% \\ \cline{2-7} 
\cellcolor[HTML]{FFFFFF} &
  Expand\_key\_type\_A &
  \multicolumn{1}{c|}{95.26\%} &
  81.23\% &
  100\% &
  \multicolumn{1}{c|}{98.16\%} &
  83.16\% \\ \cline{2-7} 
\cellcolor[HTML]{FFFFFF} &
  Expand\_key\_type\_B &
  \multicolumn{1}{c|}{100\%} &
  81.35\% &
  100\% &
  \multicolumn{1}{c|}{100\%} &
  81.44\% \\ \cline{2-7} 
\cellcolor[HTML]{FFFFFF} &
  Expand\_key\_type\_C &
  \multicolumn{1}{c|}{100\%} &
  81.20\% &
  100\% &
  \multicolumn{1}{c|}{100\%} &
  82.35\% \\ \cline{2-7} 
\cellcolor[HTML]{FFFFFF} &
  Expand\_key\_type\_D &
  \multicolumn{1}{c|}{100\%} &
  96.25\% &
  100\% &
  \multicolumn{1}{c|}{100\%} &
  96.25\% \\ \cline{2-7} 
\multirow{-6}{*}{\cellcolor[HTML]{FFFFFF}\textbf{AES-192\btrs}} &
  final\_round &
  \multicolumn{1}{c|}{100\%} &
  80.23\% &
  100\% &
  \multicolumn{1}{c|}{100\%} &
  81.60\% \\ \hline
\cellcolor[HTML]{FFFFFF} &
  des3\_top &
  \multicolumn{1}{c|}{99.82\%} &
  93.03\% &
  100\% &
  \multicolumn{1}{c|}{99.82\%} &
  93.03\% \\ \cline{2-7} 
\cellcolor[HTML]{FFFFFF} &
  CRP &
  \multicolumn{1}{c|}{99.34\%} &
  99.34\% &
  100\% &
  \multicolumn{1}{c|}{99.34\%} &
  99.34\% \\ \cline{2-7} 
\multirow{-3}{*}{\cellcolor[HTML]{FFFFFF}\textbf{DES3\btrs}} &
  key\_sel3 &
  \multicolumn{1}{c|}{82.74\%} &
  82.72\% &
  100\% &
  \multicolumn{1}{c|}{82.78\%} &
  82.75\% \\ \hline
\cellcolor[HTML]{FFFFFF} &
  gps\_top &
  \multicolumn{1}{c|}{100\%} &
  85.34\% &
  100\% &
  \multicolumn{1}{c|}{100\%} &
  83.67\% \\ \cline{2-7} 
\cellcolor[HTML]{FFFFFF} &
  AES\_192\_top &
  \multicolumn{1}{c|}{97.94\%} &
  81.25\% &
  99.56\% &
  \multicolumn{1}{c|}{97.52\%} &
  85.50\% \\ \cline{2-7} 
\cellcolor[HTML]{FFFFFF} &
  Expand\_key\_type\_A &
  \multicolumn{1}{c|}{95.26\%} &
  81.23\% &
  100\% &
  \multicolumn{1}{c|}{98.16\%} &
  83.16\% \\ \cline{2-7} 
\cellcolor[HTML]{FFFFFF} &
  Expand\_key\_type\_B &
  \multicolumn{1}{c|}{100\%} &
  81.35\% &
  100\% &
  \multicolumn{1}{c|}{100\%} &
  81.44\% \\ \cline{2-7} 
\cellcolor[HTML]{FFFFFF} &
  Expand\_key\_type\_C &
  \multicolumn{1}{c|}{100\%} &
  81.20\% &
  100\% &
  \multicolumn{1}{c|}{100\%} &
  82.35\% \\ \cline{2-7} 
\cellcolor[HTML]{FFFFFF} &
  Expand\_key\_type\_D &
  \multicolumn{1}{c|}{100\%} &
  96.25\% &
  100\% &
  \multicolumn{1}{c|}{100\%} &
  96.25\% \\ \cline{2-7} 
\cellcolor[HTML]{FFFFFF} &
  final\_round &
  \multicolumn{1}{c|}{100\%} &
  80.23\% &
  100\% &
  \multicolumn{1}{c|}{100\%} &
  81.60\% \\ \cline{2-7} 
\cellcolor[HTML]{FFFFFF} &
  pcode &
  \multicolumn{1}{c|}{94.62\%} &
  80.86\% &
  97.01\% &
  \multicolumn{1}{c|}{94.78\%} &
  82.54\% \\ \cline{2-7} 
\multirow{-9}{*}{\cellcolor[HTML]{FFFFFF}\textbf{GPS\btrs}} &
  cacode &
  \multicolumn{1}{c|}{100\%} &
  84.75\% &
  99\% &
  \multicolumn{1}{c|}{98.61\%} &
  84.75\% \\ \hline
\cellcolor[HTML]{FFFFFF} &
  fir\_top &
  \multicolumn{1}{c|}{90.58\%} &
  87.56\% &
  94\% &
  \multicolumn{1}{c|}{90.60\%} &
  87.89\% \\ \cline{2-7} 
\cellcolor[HTML]{FFFFFF} &
  fir\_filter\_block\_left &
  \multicolumn{1}{c|}{96.89\%} &
  83.56\% &
  89.23\% &
  \multicolumn{1}{c|}{95.54\%} &
  83.73\% \\ \cline{2-7} 
\cellcolor[HTML]{FFFFFF} &
  fir\_filter\_block\_left\_multiply block &
  \multicolumn{1}{c|}{92.85\%} &
  71.65\% &
  88.94\% &
  \multicolumn{1}{c|}{90.23\%} &
  71.78\% \\ \cline{2-7} 
\cellcolor[HTML]{FFFFFF} &
  fir\_filter\_block\_right &
  \multicolumn{1}{c|}{94.30\%} &
  86.70\% &
  96.24\% &
  \multicolumn{1}{c|}{94.35\%} &
  86.82\% \\ \cline{2-7} 
\multirow{-5}{*}{\cellcolor[HTML]{FFFFFF}\textbf{FIR\btrs}} &
  fir\_filter\_block\_right\_multiply block &
  \multicolumn{1}{c|}{88.56\%} &
  72.29\% &
  95.71\% &
  \multicolumn{1}{c|}{87.25\%} &
  72.29\% \\ \hline
\cellcolor[HTML]{FFFFFF} &
  iir\_top &
  \multicolumn{1}{c|}{91.78\%} &
  88.54\% &
  98.65\% &
  \multicolumn{1}{c|}{92.62\%} &
  89.34\% \\ \cline{2-7} 
\cellcolor[HTML]{FFFFFF} &
  iir\_filter\_block\_left &
  \multicolumn{1}{c|}{94.65\%} &
  80.03\% &
  100\% &
  \multicolumn{1}{c|}{94.65\%} &
  80.03\% \\ \cline{2-7} 
\cellcolor[HTML]{FFFFFF} &
  iir\_filter\_block\_left\_multiply block &
  \multicolumn{1}{c|}{89.02\%} &
  75.18\% &
  99\% &
  \multicolumn{1}{c|}{89.27\%} &
  75.58\% \\ \cline{2-7} 
\cellcolor[HTML]{FFFFFF} &
  iir\_filter\_block\_right &
  \multicolumn{1}{c|}{87.89\%} &
  72.76\% &
  98\% &
  \multicolumn{1}{c|}{88.24\%} &
  73.32\% \\ \cline{2-7} 
\multirow{-5}{*}{\cellcolor[HTML]{FFFFFF}\textbf{IIR\btrs}} &
  iir\_filter\_block\_right\_multiply block &
  \multicolumn{1}{c|}{90.05\%} &
  88.82\% &
  100\% &
  \multicolumn{1}{c|}{90.05\%} &
  88.82\% \\ \hline
 &
  bus\_monitor &
  \multicolumn{1}{c|}{91.20\%} &
  81.40\% &
  98.20\% &
  \multicolumn{1}{c|}{91.20\%} &
  81.40\% \\ \cline{2-7} 
 &
  controller module &
  \multicolumn{1}{c|}{92.65\%} &
  84.40\% &
  99.54\% &
  \multicolumn{1}{c|}{92.65\%} &
  84.40\% \\ \cline{2-7} 
\multirow{-3}{*}{\textbf{i2c\bstars}} &
  target module &
  \multicolumn{1}{c|}{90.47\%} &
  87.40\% &
  100\% &
  \multicolumn{1}{c|}{90.47\%} &
  87.40\% \\ \hline
 &
  adc\_ctrl\_fsm &
  \multicolumn{1}{c|}{85.58\%} &
  79.63\% &
  93.83\% &
  \multicolumn{1}{c|}{85.02\%} &
  79.18\% \\ \cline{2-7} 
 &
  adc\_ctrl\_intr &
  \multicolumn{1}{c|}{90.02\%} &
  82.08\% &
  94.86\% &
  \multicolumn{1}{c|}{89.58\%} &
  82.72\% \\ \cline{2-7} 
\multirow{-3}{*}{\textbf{adc\_ctrl\bstars}} &
  adc\_ctrl\_core &
  \multicolumn{1}{c|}{83.52\%} &
  79.89\% &
  96.84\% &
  \multicolumn{1}{c|}{83.76\%} &
  80.45\% \\ \hline
 &
  keccak\_round &
  \multicolumn{1}{c|}{90.52\%} &
  85.33\% &
  99.56\% &
  \multicolumn{1}{c|}{90.52\%} &
  85.50\% \\ \cline{2-7} 
 &
  sha3\_top &
  \multicolumn{1}{c|}{86.57\%} &
  78.23\% &
  95.42\% &
  \multicolumn{1}{c|}{87.16\%} &
  79.45\% \\ \cline{2-7} 
 &
  kmac core &
  \multicolumn{1}{c|}{87.54\%} &
  81.35\% &
  97.45\% &
  \multicolumn{1}{c|}{89.43\%} &
  81.44\% \\ \cline{2-7} 
\multirow{-4}{*}{\textbf{kmac\bstars}} &
  kmac\_entropy &
  \multicolumn{1}{c|}{92.76\%} &
  81.20\% &
  94.72\% &
  \multicolumn{1}{c|}{93.76\%} &
  82.35\% \\ \hline
\end{tabular}%
}
\footnotesize
{\vspace{0.75mm}\\ \btrs denotes designs from CEP SoC. \hspace{0.5em} \bstars denotes designs from OpenTitan SoC.}
\end{table*}

\subsection{Coverage Analysis at Sub-module level}
\label{appendix:cov_submodule}

Coverage analysis at the sub-module level involves assessing the checker, stimuli and formal coverage generated using Cadence JasperGold integrated with \lasa~framework. This approach enhances the overall verification efforts, allowing for a more granular exploration of design behavior at sub-module level and ensuring greater accuracy in identifying potential issues. Additionally, this approach also enhances scalability by enabling targeted testing of sub-modules, which can be independently verified and then aggregated for comprehensive verification of the overall design.

Table \ref{tab:cov_sub} presents the coverage metrics at the sub-module level for various IPs from the OpenTitan and CEP SoC benchmarks. The results highlight high coverage values for the generated properties using the \lasa~framework, demonstrating the effectiveness of the proposed approach.

\end{document}